\documentclass[12pt]{iopart}
\usepackage{graphics}
\usepackage{graphicx}
\usepackage[sort&compress,numbers]{natbib}

\begin{document}

\title[Electronic Structure of In-Plane Heterojunctions h-BN/graphene Nanoribbons]{A DFT Study on the Electronic Structure of In-Plane Heterojunctions of Graphene and Hexagonal Boron Nitride Nanoribbons}

\author{Ramiro M. dos Santos$^1$, William F. Giozza$^2$, Rafael T. de Sousa J\'unior$^2$, Dem\'etrio A. da Silva Filho$^1$, Renato B. Santos$^3$, and Luiz A. Ribeiro J\'unior$^{1,*}$}

\address{$^{1}$ \quad Institute of Physics, University of Bras\'ilia, Bras\'ilia, 70910-900, Brazil}
\address{$^{2}$ \quad Department of Electrical Engineering, University of Bras\'{i}lia 70919-970, Brazil}
\address{$^{3}$ \quad Federal Institute for Education, Science, and Technology Baiano, Senhor do Bonfim, Bahia, 48.970-000, Brazil}
\ead{ribeirojr@unb.br}
\vspace{10pt}
\begin{indented}
\item[]December 2020
\end{indented}

\begin{abstract}
The structural similarity between hexagonal boron nitride (h-BN) and graphene nanoribbons allows for the formation of heterojunctions with small chain stress. The combination of the insulation nature of the former and the quasi-metallic property of the latter makes this kind of heterostructure particularly interesting for flat optoelectronics. Recently, it was experimentally demonstrated that the shapes of the graphene and h-BN domains can be controlled precisely, and sharp graphene/h-BN interfaces can be created. Here, we investigated the electronic and structural properties of graphene (h-BN) nanoribbon domains of different sizes sandwiched between h-BN (graphene) nanoribbons forming in-plane heterojunctions. Different domain sizes for the zigzag termination were studied. Results showed that the charge density is localized in the edge of the heterojunctions, regardless of the domain size. These materials presented a metallic nature with a ferromagnetic behavior, which can be useful for magnetic applications at the nanoscale.  
\end{abstract}

\section{Introduction}

Two-dimensional nanostructured materials are promising cost-efficiency solutions for developing novel flat optoelectronic applications \cite{sun2016optical,gupta2015recent,zhang2016van,yu2020two}. Their controllable size allows for electronic wave function confinement, which is desirable for design semiconducting devices \cite{zhang2016van}. Among the materials that share this property, hexagonal boron nitride (h-BN) \cite{bechelany2008preparation,han2008structure,warner2010atomic,zhang2015two} and graphene \cite{novoselov2004electric,geim2007rise,geim2009graphene} sheets stand out. Nanoribbons are obtained by extracting a strip of the corresponding material, and they present a quasi-one-dimensional nature \cite{son2006energy,han2007energy,son2006half}. Even after producing the graphene or h-BN nanoribbon, the lattice parameter of the derived system differs only by 0.03 \r{A} concerning the original one \cite{bokai2020hybrid}. Despite the structural similarities between h-BN and graphene, it is well known that their electronic nature is substantially different \cite{sachs2011adhesion}.

Recently, several experimental \cite{wang2020towards,liu2013plane,yang2013epitaxial,song2010large,GONZALEZORTIZ2020100107,beniwal2017graphene,wu2012nitrogen,maeda2017orientation} and theoretical \cite{leon2019interface,brugger2009comparison,kaloni2012electronic,giovannetti2007substrate,slawinska2010energy,dos2019defective,dos2019electronic,zhang2015two} studies were carried out to propose routes for the precise control of the graphene and h-BN bandgaps. Among them, the substitutional doping and the reshaping of their lattice structures have been widely used \cite{doi:10.1021/jp402297n,chen2018carbon,bokdam2011electrostatic,doi:10.1021/acs.nanolett.6b03709,doi.org/10.1002/asia.201500450,enyashin2011graphene,hirsch2010era,gomes2013stability,zhao2013local}. It was experimentally demonstrated the creation of 2D in-plane graphene/h-BN heterojunctions with controlled domain sizes by using lithography patterning and sequential CVD growth steps \cite{liu2013plane}. Through this technique, the shapes of graphene and h-BN domains were controlled precisely, and sharp graphene/h-BN interfaces were created. Other materials such as MoS$_2$, AlN, and GaN have some structural properties similar to graphene \cite{splendiani2010emerging,lopez2013ultrasensitive}. However, monolayer graphene and h-BN have a lattice mismatch of only about 1.5 \% \cite{C7RA00260B}. Moreover, in the h-BN monolayer, the difference in grid energy between nitrogen atoms and boron atoms leads to a broad bandgap of about 5.9 eV, which can help to open the bandgap of graphene by producing a heterojunction between them \cite{C7RA00260B}. 

Herein, motivated by the recent achievements in the synthesis of in-plane h-BN/graphene heterojunctions \cite{liu2013plane,wang2020towards}, we used density functional theory (DFT) calculations to study the structural and electronic properties of zigzag and armchair graphene (h-BN) nanoribbons containing h-BN (graphene) domains of different sizes. Our findings revealed that the in-plane heterojunctions of h-BN/graphene nanoribbons present a ferromagnetic behavior, which can be useful for magnetic applications at the nanoscale. 

\section{Details of Modeling}

The DFT calculations of the electronic and structural properties of the in-plane heterojunctions h-BN/graphene Nanoribbons were performed using the SIESTA code \cite{soler2002siesta}. These calculations were conducted within the framework of generalized gradient approximation (GGA) with localized basis sets \cite{PhysRevLett.77.3865}, and the exchange-correlation functional was used based on the Perdew–Burke–Ernzerhof (PBE) scheme \cite{PhysRevLett.77.3865,PhysRevLett.80.891}. To treat the electron core interaction, we used the Troullier–Martins norm-conserving pseudopotentials \cite{PhysRevB.64.235111}. The polarization effects were included, and the Kohn–Sham orbitals are expanded with double-$\zeta$ basis \cite{PhysRevB.64.235111}. The structural relaxation of all model h-BN/graphene heterojunctions studied here is carried out until the force on each atom is less than $10^{-3}$ eV/\r{A} and the total energy change between each self-consistency step achieves a value less or equal to $10^{-5}$ eV. The Brillouin zone is sampled by a fine $15\times 15\times 1$ grid and, to determine the self-consistent charge density we use a mesh cutoff of 200 Ry. A supercell geometry was adopted with a vacuum distance of 30 \r{A} to avoid interaction among each structure and its images. 

\begin{figure}[!htb]
    \centering
    \includegraphics[width=0.9\linewidth]{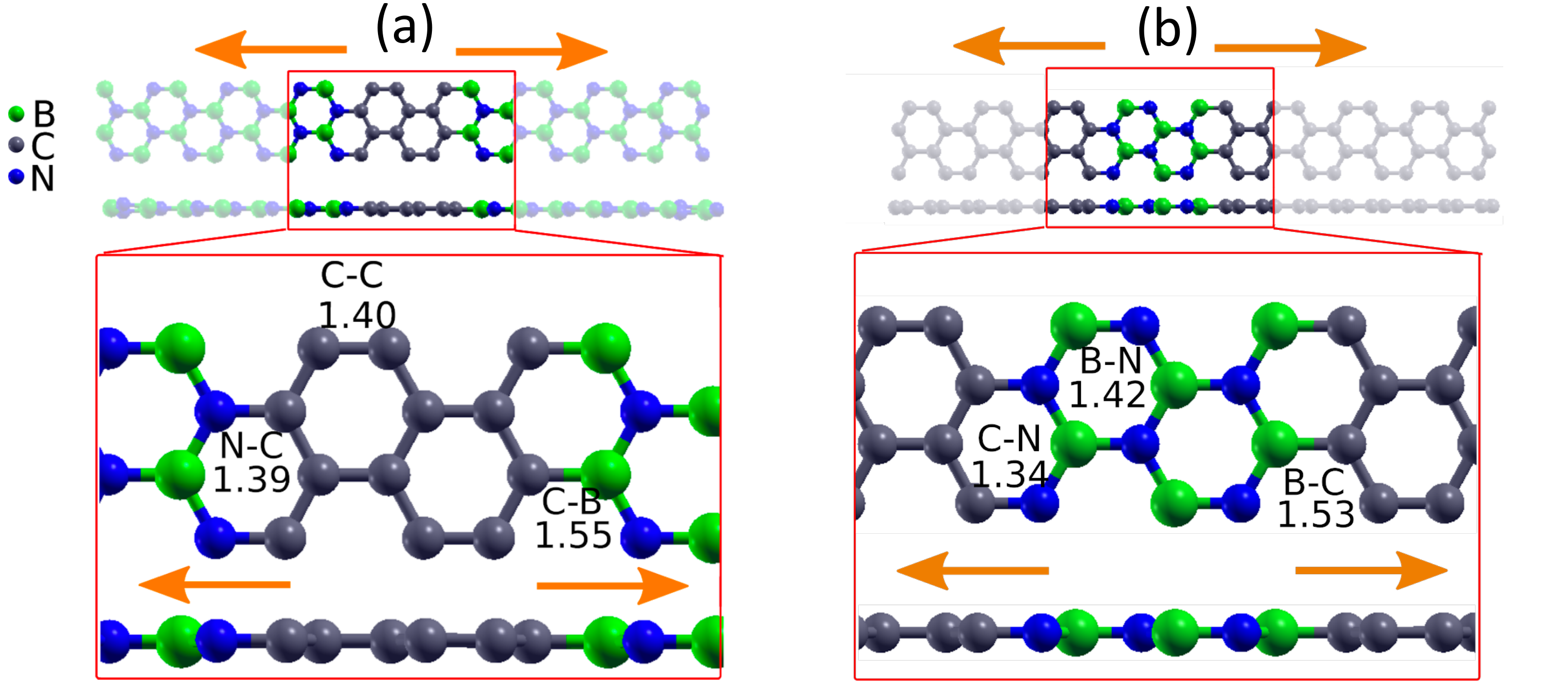}
    \caption{Diagrammatic representation of (a) $n$-ZZGBNNRs and (b) $m$-ZZGBNNRs.}
    \label{fig:systems}
\end{figure}

Figure \ref{fig:systems} illustrates the model graphene/h-BN heterojunctions studied here. Figure \ref{fig:systems}(a) represents the heterojunctions in which a zigzag h-BN nanoribbon contains a central graphene domain (named from now on as $n$-ZZGBNNRs), and Figure \ref{fig:systems}(b) depicts the cases where a graphene nanoribbon is endowed with an h-BN domain (named from now on as $m$-ZZGBNNRs). The indexes $n$ and $m$ denote the number of zigzag lines in the graphene and h-BN domains, respectively. The number of zigzag lines, which stands the domain size, varied from 1 up to 12. The inset panels in Figure \ref{fig:systems} present the C-C, B-C, and B-N bond lengths for the ground state structures of each case.     

\section{Results}

\begin{figure}[!htb]
    \centering
    \includegraphics[width=0.9\linewidth]{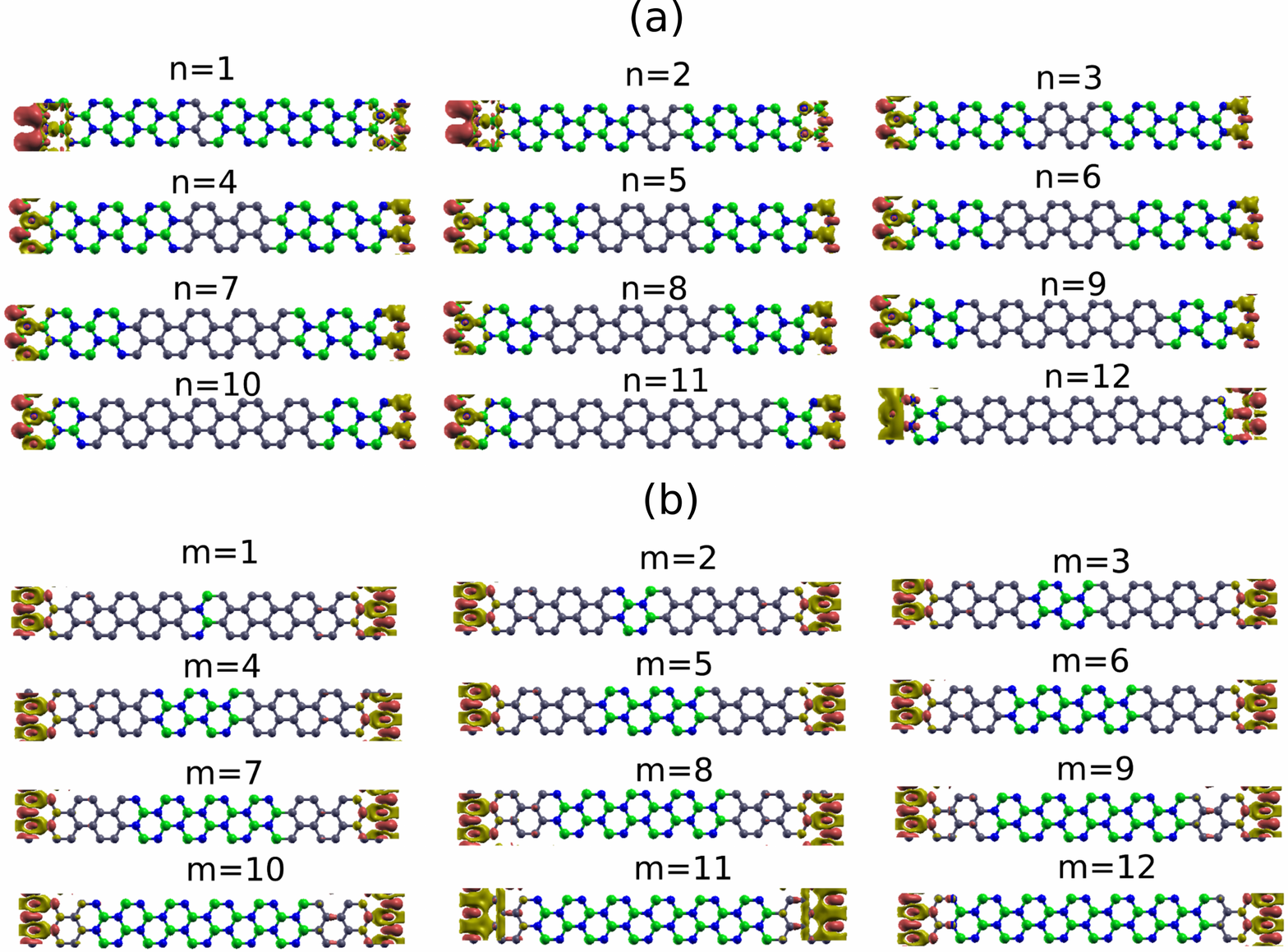}
    \caption{Schematic representation of the total charge density, that was obtained from the spin density difference $\rho(\uparrow)-\rho(\downarrow)$.}
    \label{fig:charge}
\end{figure}

We begin our discussions by presenting the total charge density for all the model heterojunctions studied here. Figure \ref{fig:charge}(a) and \ref{fig:charge}(b) illustrate the cases $n$-ZZGBNNRs and $m$-ZZGBNNRs, respectively. As a general trend, one can note in these figures that the charge density is localized at the edges of the nanoribbons, regardless of the domain size. In Figure \ref{fig:charge}(a) one can see that the left edges of the nanoribbons possess more charge concentration than the right ones. This signature for the charge distribution is related to the distinct atomic configurations at the edges. On the left edge, there are boron atoms with only one electron occupying the p-orbital, whereas the right edge is dominated by nitrogen atoms with three electrons occupying the p-orbital. In this sense, the left edge has more a pronounced electronegativity, which contributes to the symmetry breaking in the charge distribution pattern illustrated in Figure \ref{fig:charge}. On the other hand, in the $m$-ZZGBNNR cases (Figure \ref{fig:charge}(b)), we can realize that the charge density is symmetrically distributed on the ribbon's edges once their termination is equal, i.e., just carbon atoms with the same electronegativity are present.  

\begin{figure}[!htb]
    \centering
    \includegraphics[width=0.9\linewidth]{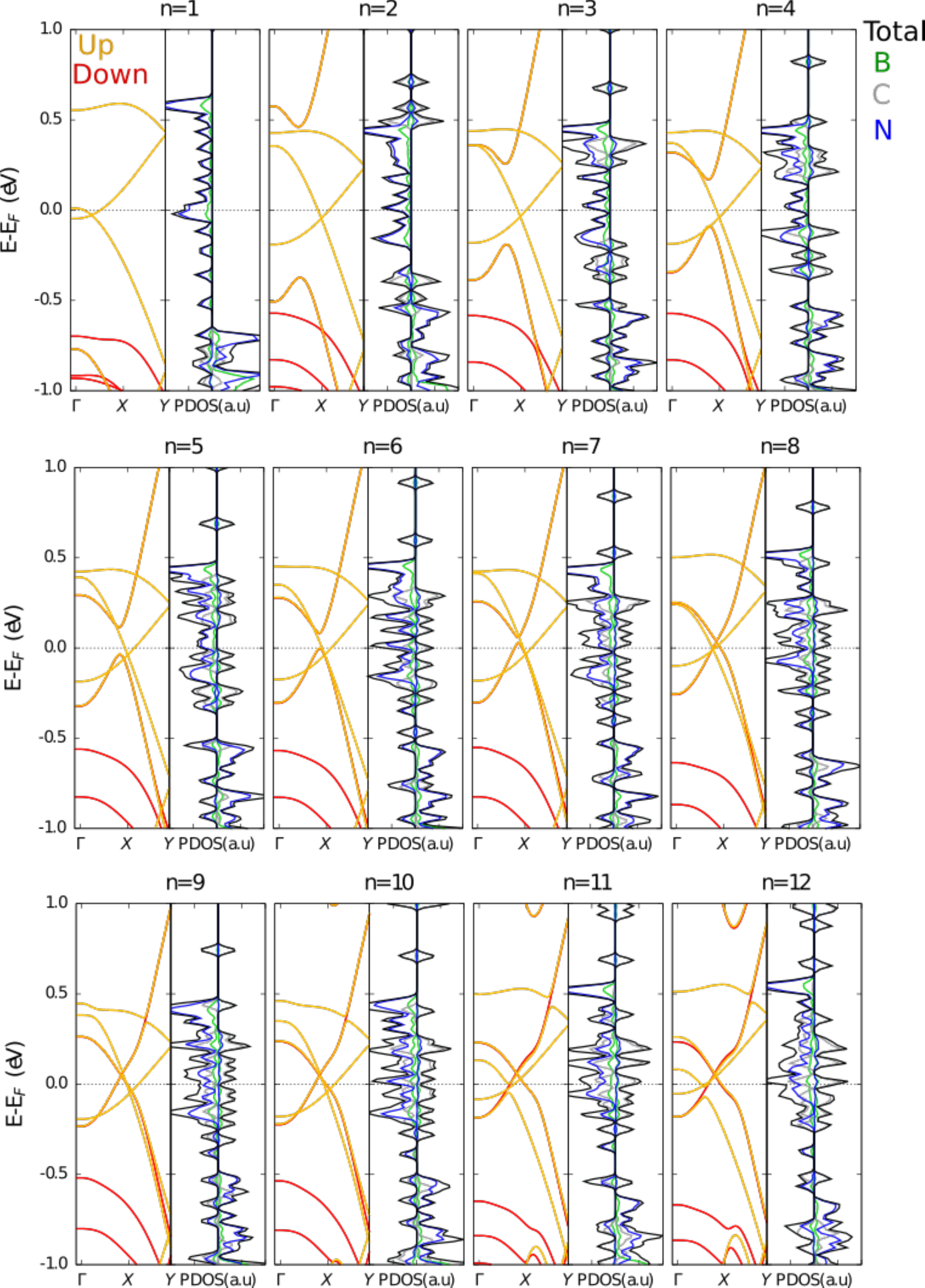}
    \caption{Electronic band structure for the $n$-ZZGBNNRs related to the cases presented in Figure \ref{fig:charge}(a).}
    \label{fig:bgn}
\end{figure}

\begin{figure}[!htb]
    \centering
    \includegraphics[width=0.9\linewidth]{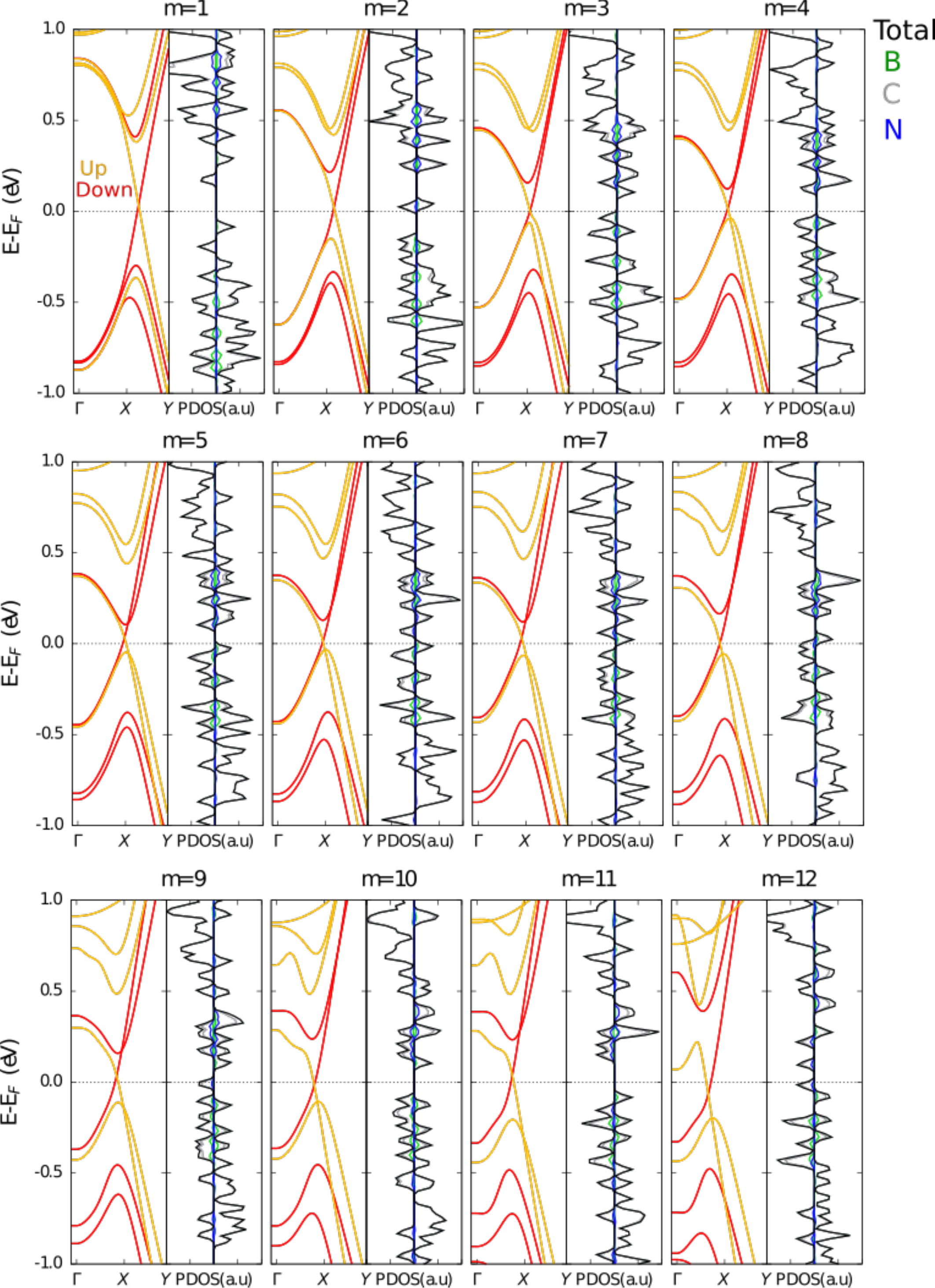}
    \caption{Electronic band structure for the $m$-ZZGBNNRs related to the cases presented in Figure \ref{fig:charge}(b).}
    \label{fig:bgm}
\end{figure}

Figures \ref{fig:bgn} and \ref{fig:bgm} show the band structures for the $n$-ZZGBNNRs and $m$-ZZGBNNRs systems, respectively, for the cases presented in Figure \ref{fig:charge}. The related partial density of states (PDOS) is depicted alongside the band structures. In these figures, one can note the clear symmetry breaking between the two spin channels --- spin-up (orange lines) and spin-down (red lines) --- that cross the Fermi level. Such a bandgap signature suggests a strong ferromagnetic behavior. This symmetry breaking in the energy levels is a consequence of the pattern for the charge density localization presented in Figure \ref{fig:charge}(a). Importantly, this ferromagnetic behavior was not predicted in other theoretical studies by using tight-binding and plane-wave DFT methods \cite{leon2019interface}. The PDOS shows that the major contribution to the intragap levels comes from boron and nitrogen atoms. These levels belong to the chemical species that compose the ribbon's edges, where the net charge is concentrated due to the presence of dangling bonds. In this sense, the edge states are crucial in characterizing the electronic transport and the magnetic moment of these h-BN/graphene heterojunctions. In Figure \ref{fig:bgm}, we note that the energy levels near the Fermi level for the $m$-ZZGBNNRs band structures present a higher degree of symmetry when contrasted to the $n$-ZZGBNNR cases. As expected, the PDOS shows a major contribution of carbons atoms for these levels. The overall ferromagnetic behavior is smaller in the $m$-ZZGBNNR cases due to the degree of symmetry showed by the energy levels that crossed the Fermi level.           

\section{Conclusion} 
In summary, the electronic and structural properties of in-plane heterojunctions of h-BN/graphene nanoribbons were numerically studied using DFT calculations. The model heterojunctions studied were composed of a zigzag h-BN nanoribbon containing a central graphene domain (named $n$-ZZGBNNRs) and a zigzag graphene nanoribbon endowed with an h-BN domain (named $m$-ZZGBNNRs). The indexes $n$ and $m$ denote the number of zigzag lines in the graphene and h-BN domains, respectively. These indexes defined the domain size. Our computational protocol was based on performing the DFT calculations in heterojunctions of different domain sizes. Results showed that the charge density is localized in the edge of the heterojunctions, regardless of the domain size. As a consequence, these heterojunctions presented a ferromagnetic behavior, which can be interesting for magnetic applications in flat optoelectronics. 

\section{Acknowledgments}
The authors gratefully acknowledge the financial support from Brazilian Research Councils CNPq, CAPES, and FAPDF and CENAPAD-SP for providing the computational facilities. W.F.G. gratefully acknowledges the financial support from FAP-DF grant $0193.0000248/2019-32$. L.A.R.J. gratefully acknowledges the financial support from CNPq grant $302236/2018-0$. R.T.S.J. gratefully acknowledges, respectively, the financial support from CNPq grant $465741/2014-2$, CAPES grants $88887.144009/2017-00$, and FAP-DF grants $0193.001366/2016$ and $0193.001365/2016$. L.A.R.J. gratefully acknowledges the financial support from DPI/DIRPE/UnB (Edital DPI/DPG $03/2020$) grant $23106.057541/2020-89$ and from IFD/UnB (Edital $01/2020$) grant $23106.090790/2020-86$. D.A.S.F acknowledges the financial support from the Edital DPI - UnB N. $04/2019$, from CNPq (grants $305975/2019-6$ and $420836/2018-7$) and FAP-DF grants $193.001.596/2017$ and $193.001.284/2016$.

\bibliographystyle{iopart-num}
\bibliography{references}

\end{document}